\documentclass[dvips]{arxstspdf}
\usepackage{graphics,flushend}

\volume{25}
\issue{2}
\pubyear{2010}
\firstpage{202}
\lastpage{216}
\doi{10.1214/10-STS332}



\begin{document}
\begin{frontmatter}

\title{Continual Reassessment and Related Dose-Finding Designs}
\runtitle{Continual Reassessment and Related Dose-Finding Designs}

\begin{aug}
\author[a]{\fnms{John} \snm{O'Quigley}\corref{}\ead[label=e1]{jmoquigley@gmail.com}}
and
\author[b]{\fnms{Mark} \snm{Conaway}\ead[label=e2]{mconaway@virginia.edu}}
\runauthor{J. O'Quigley and M. Conaway}

\affiliation{Universit\'e Paris VI and University of Virginia}

\address[a]{John O'Quigley is Professor, Inserm, Universit\'e
Paris VI, Place Jussieu, 75005 Paris, France \printead{e1}.}
\address[b]{Mark Conaway is Professor, Division of Biostatistics,
Department of Public Health Sciences, University of Virginia,
Charlottesville, VA 22908, USA \printead{e2}.}

\end{aug}

%
\begin{abstract}
During the last twenty years there have been considerable
methodological developments in the design and analysis of Phase 1,
Phase 2 and Phase 1/2
dose-finding studies. Many of these developments are related to the
continual reassessment method (CRM), first introduced by
O'Quigley, Pepe and Fisher (\citeyear{QPF1990}).
CRM models have proven themselves
to be of practical use and, in this discussion, we
investigate the basic approach, some connections to other methods, some
generalizations, as well
as further applications of the model. We obtain some new results which
can provide guidance in
practice.
\end{abstract}

%

\end{frontmatter}

\section{Introduction}\label{sec1}

The continual reassessment method (CRM) was introduced by O'Quigley,
Pepe and Fisher (\citeyear{QPF1990}) as
a design with which to carry out and analyze dose-finding studies in
oncology. The purpose of these
studies, usually referred to as Phase 1 trials of a new therapeutic
agent, is to estimate
the maximum tolerated dose (MTD) to be used in Phase 2 and Phase 3 trials.
O'Quigley, Pepe and Fisher (\citeyear{QPF1990}) pointed out that
standard methods in use then, and still in use
now, fail to address the basic ethical requirements of experimentation
with human
subjects. Given the unknown or
poorly understood relationship between dose and the probability of
undesirable side effects (toxicity),
it is inevitable, during experimentation,
that some patients will be treated at too toxic doses and some patients
will be treated
at doses too low to have any real chance of procuring benefit. Aside
from being inefficient,
the case against the standard designs
is that more patients than necessary are treated in this way, either at
too toxic a dose or, more
usually, at too low a dose to provide therapeutic benefit.

The rationale of the CRM is to concentrate as many patients as we can
on doses at, or close to,
the MTD. Doing so can provide an efficient estimate of the MTD while
maximizing the number of patients in the study treated at doses with
potential therapeutic benefit but without undue risk of toxicity.
A drawback of concentrating patients to a small number of dose levels,
at and around the MTD, is that the overall dose-toxicity
curve itself may be difficult to estimate. In practice, this tends not
to be a serious drawback, since
estimation of the entire dose-toxicity curve is rarely the goal of a
dose-finding clinical trial.

Phase 1 trials evaluating the toxicity of single agents are becoming
less common, giving way to more complex studies involving multiple
agents at various doses, heterogeneous groups of patients, and
evaluations of both toxicity and efficacy. The standard methods are
ill-equipped to handle these more complex situations, and here, we will
discuss developments of the CRM and related methods for tackling
various problems which arise in the context of dose finding. Whereas
the standard method, even for the simplest situation fails to perform
adequately, model based designs, while offering greatly improved
performance for the simplest case, allow us to take on board those more
involved situations that arise in practice (Braun,
\citeyear{B2002};
Faries, \citeyear{F1994}; Goodman, Zahurak and Piantadosi,
\citeyear{GZP1995}; Legedeza and Ibrahim, \citeyear{LI2002}; Mahmood,
\citeyear{M2001}; O'Quigley, \citeyear{Q2002a}; O'Quigley and Paoletti,
\citeyear{QP2003};
O'Quigley and Reiner, \citeyear{QR1998};\break O'Quigley, Shen and Gamst, \citeyear{QSG1999}; Piantadosi
and Liu, \citeyear{PL1996}).

We begin with the definitions and notation used in Phase 1 trials and
an overview of the CRM as originally proposed by
O'Quigley, Pepe and Fisher (\citeyear{QPF1990}). The next two sections
outline Bayesian and likelihood-based inference for the CRM, providing
results for the small-sample and large-sample properties of the method.
Section \ref{sec5} gives extensions of the method and discusses modifications of
the basic design. Section \ref{sec6} presents related designs, again for the
case of a single outcome whereas
Section \ref{sec7} considers two outcomes, one positive and one negative and
describes the goal of locating the most successful dose (MSD). The
article concludes with a discussion of future directions in the study
of model-based methods for dose-finding studies.

\subsection{Doses, DLT, MTD and the MSD}\label{sec1.1}


Traditional thinking in the area of cytotoxic anti-cancer treatments is
to give as
strong a treatment as we can without incurring too much toxicity.
For the great majority of new cancer treatments---recent advances
in immunotherapy being possible exceptions---we consider that
increases in dose correspond to increases in both the number of
patients who will experience toxic side effects as well as the
numbers who may benefit from treatment. If we observe
a complete absence of toxic side effects, then we would not
anticipate observing any therapeutic effect, either for those
patients in the study or for future patients. The Phase I trial
then has for its goal the determination of some dose having an
``acceptable'' rate of toxicity. While it is true that the
essential goal of the study is to improve treatment for future
patients, ethical concerns dictate that we give the best possible
treatment to the patients participating in the Phase I study
itself. The highest dose level at which patients can be treated
and where the rate of toxicity is deemed to be still acceptable is
known as the MTD (maximal tolerated dose).

On an individual level we can imagine being able to increase the dose without
encountering the toxic effect of interest. At some threshold the individual
will suffer a toxicity. An assumed model is the following: at this
threshold the individual
suffers a toxicity and, for all higher doses, the individual would also
have encountered
a toxicity. Such a model is reasonable for most situations and widely
assumed. It remains nonetheless
a model and might be brought under scrutiny in particular cases. The
model stipulates that for all levels below the
threshold, the individual would not suffer any toxicity and we call the
threshold itself the
individual's own maximum tolerated dose (MTD).
A dose-limiting toxicity (DLT) curve for the individual would be a
$(0,1)$ step function, the value 0
indicating no toxicity and the value 1 a toxicity. Thus, in the case of
an individual, the $(0,1)$
step function for the DLT coincides with that for the MTD.

Any population of interest can be viewed as being composed of
individuals each having
their own particular MTD. Corresponding to each individual MTD we have
a $(0,1)$ step function for
the individual's DLT. Over some set or population of individuals, the
sum of the
DLT curves at any dose equates to the probability of toxicity at that
same dose.
For a population we fix some percentile so that, $100\times\theta \%
$ say, have their own threshold at or
below this level. The term MTD is often used somewhat loosely, and not
always well defined. The more precise
definition given in terms of a percentile involves $\theta.$
Different values of $\theta$ would correspond to different definitions
of the MTD. The values 0.2, 0.25 and 0.33 are quite common in practice.

When information on efficacy, possibly through surrogate measures or
otherwise through
some measure of response, is available in a timely way, then it makes
sense to make use of such information.
In the HIV setting, there have been
attempts to simultaneously address the problems of both toxicity and
efficacy. The goal then becomes not one of finding the maximum tolerated
dose but, rather, one of finding the MSD (most successful dose), that
is, that
dose where the probability of treatment failure, be it due to excessive
toxicity or to insufficient evidence of treatment efficacy, is a
minimum. The CRM
can be readily adapted to address these kinds of questions (O'Quigley,
Hughes and Fenton, \citeyear{QHF2001};
Zohar and O'Quigley, \citeyear{ZQ2006a}).

\subsection{Notation}

We assume that we have available $k$ doses; $d_1,\ldots,\break d_k$,
possibly multidimensional and ordered in terms of the
probabilities, $R(d_i)$, for toxicity at each of the levels, that is,
$R(d_i)< R(d_j)$ whenever $i<j$.
The MTD is denoted $d_0$ and is taken to be one of the values in the
set $\{d_1,\ldots, d_k
\}.$ It is the dose that has an associated probability of toxicity,
$R(d_0)$, as close as we can get to some target ``acceptable''
toxicity rate $\theta$. Specifically we define $d_0\in
\{d_1,\ldots, d_k \}$ such that
\begin{eqnarray}\label{disttheta}
&&| R(d_0)-\theta |\nonumber\\[-8pt]\\[-8pt]
&&\quad <| R(d_\ell
)-\theta |   , \quad \ell=1, \ldots, k;  d_\ell\ne d_0.\nonumber
\end{eqnarray}
%
The binary indicator
$Y_{j}$ takes the value $1$ in the case of a toxic response for
the $j $th entered subject ($j=1,\ldots, n$) and 0 otherwise. The
dose for the $j $th entered subject, $X_j$, is viewed as random
taking values $x_j\in\{ d_1,\ldots, d_k \}  ;j=1,\ldots,n.$ Thus we
can write
\[
\Pr (Y_j=1|X_j=x_j)=R(x_j).
\]
Little is known about $R(\cdot)$ and, given the $n$ observations,
the main goal is to identify $d_0$. Estimation of all or part
of $R(d_\ell)  ,$ $\ell=1,\ldots,k,$ is only of indirect interest
in as much
as it may help provide information on~$d_0.$

There is an extensive literature on problems similar to that just described.
The solutions to these problems, however, are mostly inapplicable in view
of ethical constraints involved in treating human subjects.
The patients included in the Phase I
design must, themselves, be treated ``optimally,'' the notion
optimal now implying for these patients a requirement to treat at
the best dose level,
taken to be the one as close as we can get to $d_0$.
We then have two statistical goals: (1) estimate
$d_0$ consistently and efficiently and, (2) during the course of
the study, concentrate as many experiments as possible around
$d_0$. Specifically, we aim to treat the $j $th included patient at the
same level we would have estimated as being $d_0$ had the study
ended after the inclusion of $j-1$ patients.


\section{Continual Reassessment Method}\label{sec2}
The continual reassessment method (CRM), proposed as a
statistical design to meet the requirements of the type of studies
described above, was introduced by O'Quigley, Pepe and Fisher
(\citeyear{QPF1990}). Many developments and innovations have followed,
the basic
method and variants having found a number of other potential
applications. Here, we reconsider the
original problem, expressed in statistical terms, since it is
this problem that forged the method. In this article we consider
the main theoretical ideas and do not dwell on precise
applications apart from for illustrative purposes.

The method begins with a
parameterized
working model for $R(x_j)$, denoted by $\psi(x_j,a)$,
for some one-parameter model $\psi(x_j,a)$ and $a$
defined on the set $\mathcal{A}$. For every $a$, $\psi(x,a)$ should be
monotone increasing in $x$ and, for any $x$, $\psi(x,a)$ should
be monotone in $a$. For every $d_i$ there exists some $a_i\in
{\mathcal{A}}$ such that $R(d_i)= \psi(d_i,a_i)$, that is, the
one-parameter model is rich enough, at each dose, to exactly
reproduce the true probability of toxicity at that dose.
There are many choices for $\psi(x,a)$, including the simple
Lehmann type shift model:
\begin{eqnarray}\label{powermodel}
&&\log\{-\log\psi(d_i,a) \} \nonumber\\[-8pt]\\[-8pt]
&&\quad = \log\{-\log\alpha_i \} + a ,\quad
i=1,\ldots,k,\nonumber
\end{eqnarray}
where $0<\alpha_1< \cdots<\alpha_k<1$ and $-\infty<a<\infty$,
having shown itself to work well in practice. This parameterization allows
for the support of the parameter $a$ to be on the whole real line and
priors such
as the normal or the logistic, having heavier tails, have been used
here. The simple
power model of O'Quigley, Pepe and Fisher (\citeyear{QPF1990})
required that support for
the parameter $a$ be restricted to the positive real line.
%

O'Quigley, Pepe and Fisher (\citeyear{QPF1990}) suggested that the
${\alpha_i}, i = 1,\ldots,k$, be chosen to reflect a priori
assumptions about the toxicity probabilities associated with each dose.
Lee and Cheung (\citeyear{LC2009}) provided a systematic approach to choosing the
initial values for the ${\alpha_i}, i = 1,\ldots,k$. Yin and Yuan
(\citeyear{YY2009}) used Bayesian model averaging to combine estimates from
different sets of initial guesses at the
${\alpha_i}, i = 1,\ldots,k$. It should again be noted that the
working model is not anticipated to represent the entire dose-toxicity
curve. It suffices that the parameterized working model be flexible
enough to allow for estimation of the dose-toxicity relationship at and
close to the MTD. This point will be developed more fully in Section \ref{sec.theprop},
which summarizes the small- and large-sample properties of the CRM.
Cheung and Chappell (\citeyear{CC2002}) investigated the operational sensitivity to
different model choices.


Once a model has been chosen and we have data in
the form of the set $\Omega_j=\{y_1,x_1,\ldots,y_{j},x_{j}\}$, the
outcomes of the first $j$ experiments, we obtain estimates ${\hat
R}(d_i)$ $ (i=1,\ldots,k)$ of the true unknown probabilities
$R(d_i)$ $ (i=1,\ldots,k)$ at the $k$ dose levels (see below).
The target dose level is that level having associated with it a
probability of toxicity as close as we can get to $\theta$. The
dose or dose level $x_j$ assigned to the $j $th included patient
is such that
\begin{eqnarray}\label{disthat}
&&|  {\hat R}(x_j)-\theta  | \nonumber\\[-8pt]\\[-8pt]
&&\quad < |
{\hat R}(d_\ell) -\theta |  ,\quad  \ell=1,\ldots,k;  d_\ell\ne
x_j.\nonumber
\end{eqnarray}
This equation should be compared to (\ref
{disttheta}). It translates the idea that the overall goal of
the study is also the goal for each included patient.
The CRM is then an iterative sequential design, the level chosen
for the $(n+1)$th patient, who is hypothetical, being also our
estimate of $d_0.$ After having included $j$ subjects, we can calculate
a posterior distribution for $a$ which we denote by $f(a,\Omega_j).$
We then induce
a posterior distribution for ${\psi}(d_i,a)$, $i=1,\ldots,k,$ from
which we can obtain summary estimates of the toxicity probabilities at each
level so that
%
\begin{eqnarray}
&&{\hat R}(d_i)\nonumber\\[-8pt]\\[-8pt]
&&\quad  = \int_{a\in\mathcal{A}} \psi(d_i,a)
f(a,\Omega_j) \,da   ,\quad i=1,\ldots,k .\nonumber
\end{eqnarray}
Using (\ref{disthat}) we can now decide which dose level to allocate
to the $(j+1)$th patient.

In the original version of the CRM, O'Quigley, Pepe and Fisher
(\citeyear{QPF1990}) used an
alternative estimate ${\tilde R}(d_i) = \psi(d_i,\mu) $, $i=1,\ldots
,k, $
where $\mu=\break \int_{a\in\mathcal{A}} a f(a,\Omega_j) \,da   .$
This was done primarily to reduce the amount of calculation required, a
consideration of less importance today. O'Quigley, Pepe and Fisher
(\citeyear{QPF1990}) completed the specification of the dose
allocation algorithm by specifying a starting dose based on a prior
specification of the dose level with probability closest to the target.


\section{Bayesian and Likelihood Inference}\label{sec3}
In order to base inference only on the likelihood it is necessary to
have the likelihood
nonmonotone so that the estimates are not on the boundary of the
parameter space. This is
accomplished by having some initial escalation scheme until the data
achieve at least one
toxicity and one nontoxicity. We can regard the data obtained via this
initial escalation scheme as, in some
sense, empirical and use them as a data-based prior to the second part
of the study. Thus, both Bayesian
and likelihood alone, can all be put under a Bayesian heading. We use
this in the following to study
different Bayesian approaches to inference.


\subsection{Likelihood-Based Dose Allocations}\label{sec3.1}
After the inclusion of the first $j$ patients, the
logarithm of the likelihood can be written as
\begin{eqnarray}\label{Loglike}
{\mathcal{L}}_j(a) & = & \sum_{\ell=1}^{j} y_\ell  \log
\psi(x_\ell,a)\nonumber\\[-8pt]\\[-8pt]
&&{} + \sum_{\ell=1}^{j} (1-y_\ell)
\log\bigl(1-\psi(x_\ell,a)\bigr),
\nonumber
\end{eqnarray}
where any terms not involving the parameter $a$ have been equated to zero.
We suppose that ${\mathcal{L}}_j (a)$
is maximized at $a=\hat{a}_j$. Once we have calculated
$\hat{a}_j$ we can next obtain an estimate of the probability of
toxicity at each dose level $d_i$ via $ {\hat
R}(d_i)=\psi(d_i,\hat{a}_j)   $ $  (i=1,\ldots,k). $
On the basis of this formula the dose to be given to the
$(j+1) $th patient, $x_{j+1}$, is determined.
Once we have estimated $a$ we can also calculate an approximate
100$(1-\alpha)\% $ confidence interval for $\psi(x_{j+1},{\hat
a}_j)$ as $(\psi_j^- ,\psi_j^+)$ where
\begin{eqnarray*}
\psi_j^- &=&\psi\bigl\{
x_{j+1},\bigl({\hat a}_j +z_{1-\alpha/2}v({\hat a}_j)^{1/2} \bigr)\bigr\}  ,\\
\psi_j^+ &=&\psi\bigl\{ x_{j+1},\bigl({\hat a}_j -z_{1-\alpha/2}v({\hat
a}_j)^{1/2} \bigr)\bigr\},
\end{eqnarray*}
where $z_\alpha$ is the $\alpha $th
percentile of a standard normal distribution and $v({\hat a}_j)$
is an estimate of the variance of ${\hat a}_j$. For the model of
(\ref{powermodel}) this turns out to be particularly simple and we can
write
\begin{eqnarray*}
v^{-1}({\hat a}_j) &=& \sum_{\ell\leq j , y_\ell=0}
\psi(x_\ell,{\hat a}_j) (\log\alpha_\ell)^2 \\
&&\hphantom{\sum_{\ell\leq j , y_\ell=0}}{}/
\bigl(1- \psi(x_\ell,{\hat a}_j)\bigr)^{2}.
\end{eqnarray*}
Although based on a misspecified model these intervals turn out
to be quite accurate, even for sample sizes as small as 16, and
thus helpful in practice (O'Quigley, \citeyear{Q1992}).

\subsection{Prior Information on the Parameter $a$}\label{sec.prior}
There are three distinct approaches which can be used in order to establish
the prior information. These are: (1) postulate some numerically
tractable and
sufficiently flexible density $g(a)$, (2) introduce
a\break pseudo-data prior which indirectly will specify $g(a)$, and (3) use
some initial
escalation scheme in a two-stage design until the first toxicity is
observed. Let us
consider these three approaches more closely.

\subsubsection*{A gamma prior for $g(a)$} For the Lehmann shift
model, on a logarithmic
scale, given that ${\mathcal{A}}=(0,\infty),$ O'Quigley, Pepe and Fisher
(\citeyear{QPF1990}) suggested,
as a natural candidate,
%
\begin{eqnarray}
g(a) =\lambda^c a^{c-1} \exp\{-(\lambda a)\}/\Gamma(c),  \nonumber\\
\eqntext{\displaystyle\Gamma(c)=
\int_0^\infty\exp(-u)u^{c-1}  \,du}
\end{eqnarray}
the gamma density with scale parameter $\lambda$ and shape parameter $c$.
%
The necessary steps in fitting a gamma
prior on the basis of the upper and lower points of our prior confidence
region have been described by Martz and Waller (\citeyear{MW1982}). For a relatively
simple set-up involving no more
than six doses and using a coding for dose (not the actual dose itself),
O'Quigley, Pepe and Fisher (\citeyear{QPF1990}) suggested that the
simple exponential prior for $a$---a special case of the gamma prior
with $c$
and $\lambda$ both equal to 1---would be satisfactory. Some authors
have appealed to this simple exponential prior in different
contexts, or more involved set-ups, and the resulting behavior of the
method can be alarming (Moller, \citeyear{M1995}).

\subsubsection*{Pseudo-data prior}
In the place of a prior expressed as a specific parametric density
function, pseudo-data priors create observations
that are weighted in accordance with our degree of belief in their plausibility.
Using pseudo-data ${y^*_{\ell}}, \ell= 1, \ldots, K$, the prior
$g(a)$ is defined from
\begin{eqnarray}\qquad g(a) &\cong&\exp \Biggl[ \sum_{\ell=1}^{j} y^*_\ell
\log \psi(x_\ell,a) \nonumber\\[-8pt]\\[-8pt]
&&\hphantom{\exp \Biggl[ }
{}+ \sum_{\ell=1}^{j} (1-y^*_{\ell})   \log
\bigl(1-\psi(x_\ell,a)\bigr)  \Biggr].\nonumber
\end{eqnarray}
The prior ``data'' can be combined with the observed data. In consequence
standard and widely available programs such as SAS or SPSS may be used
directly to calculate the posterior mode without the need for numerical
integration. The pseudo-data prior can
be used to establish our best prior guesses which will be mirrored by
the estimates
of $a$ obtained from fitting the pseudo-data alone. The imprecision
which we wish to associate
with this can be governed by a weighting coefficient $w_j$ where
$0<w_j<1.$ This
coefficient can be independent of $j$ and we would usually require that
$w_j\leq w_{j-1}.$
The posterior density is then
\begin{eqnarray}
f(a,\Omega_j) &=& A^{-1}_j \exp \{ w_j \log g(a)\nonumber\\[-8pt]\\[-8pt]
&&\hphantom{A^{-1}_j \exp \{ }
{} +
(1-w_j ) {\mathcal{L}}_j(a)  \},\nonumber
\end{eqnarray}
where $A_j=\! \int_{-\infty}^\infty\exp\{ w_j \!\log g(a) + (1-w_j )
{\mathcal{L}}_j(a) \} \,da .$
The added generality of allowing the dependence of the weights on $j$
would rarely be needed
and, in most practical situations, it suffices to take $w$ as a
constant small enough so
that the prior has no more impact than deemed necessary.

\begin{figure*}

\includegraphics{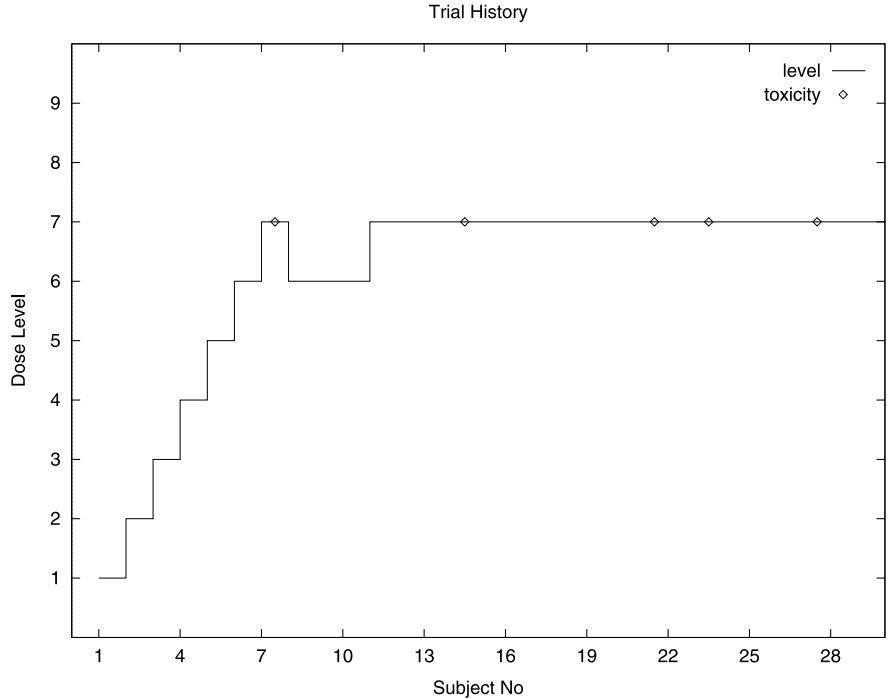}

\caption{A typical trial history using rapid early escalation; target
is level 7.}
\label{fig:trihist}
\end{figure*}

\subsubsection*{Uninformative priors}
For the model (\ref{powermodel}), O'Quigley (\citeyear{Q1992}) suggested a normal
prior having
mean zero and variance $\sigma^2$, large enough to be considered
noninformative. Such a concept can be made more precise in the
following way, at least for fixed sample designs. The mean and
mode of the prior are at zero so that, should the true
probabilities of toxicity exactly coincide with the $\alpha_i$
then, the more informative the prior the better we do, ultimately
as the prior tends to being degenerate, that is, $\sigma^2\to0$, we
obtain the correct level always. Taking some distance measure
between the distribution of our final recommendation and the
degenerate distribution putting all mass on the correct level,
this distance will increase as our uncertainty, as measured by
$\sigma^2$, increases. The curve of this distance, as a function of
$\sigma^2$, will reach an asymptotic limit, further increases in
$\sigma^2$ having a vanishing influence on the error
distribution of final recommendation. The smallest finite value
of $\sigma^2$, such that the operating characteristics are
sufficiently close to those obtained when $\sigma^2$ is infinite
(in practice very large), corresponding to a diffuse and even
improper prior, will provide the prior with the required behavior.

An uninformative prior, in the sense that it does not favor any particular
level, can be constructed readily in the light of the results of
O'Quigley (\citeyear{Q2006})
which partition the interval $[A,B]$ for the parameter $a$ into $k$ subintervals
$S_i  $ $(i=1, \ldots, k).$ If $a\in S_i  ,$ then dose level $d_i$
corresponds to the MTD.
For $k$ dose levels we simply associate
the probability mass $1/k$ to each of the $k$ subsets $S_i.$ Clearly
this approach
is readily extended to the informative case by putting priors favoring
some levels more
than others, either on the basis of clinical information or
simply out of a desire to influence the operating characteristics
in some particular way. An example for the frequent case $k=6$
would be to associate the prior 0.05 with level 1, and the values
0.19 with the other five levels. This would result in steering us
away from level 1 in favor of the other levels, unless the
accumulating data begin to weigh against our conjecture that
level 1 is unlikely to be the right level.

\subsubsection*{Data-based prior in two-stage designs}

In order to be able to maximize the log-likelihood on the
interior of the parameter space we require heterogeneity among the
responses, that is, at least one toxic and one nontoxic response
(Silvapulle, \citeyear{S1981}). Otherwise the likelihood is maximized on the
boundary of the parameter space and our estimates of $ R(d_i)$
$(i=1, \ldots, k)$ are trivially either zero, 1, or, depending
on the model we are working with, may not even be defined. In the context
of ``pure likelihood''-based designs O'Quigley and Shen (\citeyear{QS1996}) argued
for two-stage designs whereby an initial escalation scheme provided the
required heterogeneity. The experiment can be viewed as not being fully
underway until we
have some heterogeneity in the responses. These could arise in a
variety of different ways: use of a standard Up and Down
approach, use of an initial Bayesian CRM as outlined below, or use
of a design believed to be more appropriate by the investigator.
Once we have achieved heterogeneity, the model kicks in and we
continue as prescribed above (estimation--allocation). We can also consider
this initial escalation as providing empirical data. Conditional upon these
data we then proceed to the second stage.
The data obtained at the end of the first stage can be viewed as
providing an
empirical prior. In this way, all the approaches can be grouped under a
Bayesian umbrella.
The essential differences arise through the different
ways of specifying the prior.

Using empirical data to construct a prior as the first stage of a
two-stage design can afford us a
great deal of flexibility.
%
The initial exploratory escalation stage is followed
by a more refined homing in on the target. Such an idea was first
proposed by Storer (\citeyear{S1989}) in the context of the more classical Up
and Down schemes. His idea was to enable more rapid escalation in
the early part of the trial where we may be quite far from a
level at which treatment activity could be anticipated. Moller
(\citeyear{M1995}) was the first to use this idea in the context of CRM
designs. Her idea was to allow the first stage to be based on
some variant of the usual Up and Down procedures.
In the context of sequential likelihood estimation, the necessity
of an initial stage was pointed out by O'Quigley and Shen (\citeyear{QS1996})
since the likelihood equation fails to have a solution on the
interior of the parameter space unless some heterogeneity in the
responses has been observed. Their suggestion was to work with
any initial scheme, Bayesian CRM or Up and Down, and, for any
reasonable scheme, the operating characteristics appear
relatively insensitive to this choice.

%
\begin{table}[b]
\caption{Toxicity ``grades'' (severities) for trial}\label{table1}
\begin{tabular*}{7cm}{@{\extracolsep{\fill}}ll@{}}
\hline
\textbf{Severity} & \multicolumn{1}{l@{}}{\textbf{Degree of toxicity}} \\
\hline
0 & No toxicity \\
1 & Mild toxicity (non-dose-limiting) \\
2 & Nonmild toxicity (non-dose-limiting) \\
3 & Severe toxicity (non-dose-limiting) \\
4 & Dose-limiting toxicity \\
\hline
\end{tabular*}
\end{table}%
%
%

%
%
Here we describe an example of a two-stage design that has been
used in practice (see Figure \ref{fig:trihist}). There were many dose levels and the first
included patient was treated at a low level. As long as we observe
very low-grade toxicities then we escalate quickly, including only
a single patient at each level. As soon as we encounter more
serious toxicities then escalation is slowed down. Ultimately we
encounter dose-limiting toxicities at which time the second stage,
based on fitting a CRM model, comes fully into play. This is done
by integrating this information and that obtained on all the
earlier non-dose-limiting toxicities to estimate the most
appropriate dose level.
Consider the following example
which uses information on low-grade toxicities in the
first stage in order to allow rapid initial
escalation (see Table \ref{table1}).
Specifically we define a grade severity variable $S(i)$ to
be the average toxicity severity observed at dose level~$i$, that is,
the sum of the severities at that level divided by the number of
patients treated at that level.
The rule is to escalate providing $S(i)$ is less than 2.
Furthermore, once
we have included three patients at some level, then escalation to
higher levels only occurs if each cohort of three patients does not
experience dose-limiting toxicity. This scheme means that, in
practice, as long as we see only toxicities of severities coded 0
or 1, then we escalate. The first severity coded 2 necessitates a
further inclusion at this same level and, anything other than a 0
severity for this inclusion, would require yet a further inclusion
and a non-dose-limiting toxicity before being able to escalate.
This design also has the advantage that, should we be slowed down
by a severe (severity 3), albeit non-dose-limiting toxicity, we
retain the capability of picking up speed (in escalation) should
subsequent toxicities be of low degree (0 or 1). This can be
helpful in avoiding being handicapped by an outlier or an
unanticipated and possibly not drug-related toxicity arising
early in the study. Once a dose-limiting toxicity is encountered the
initial escalation stage is brought to a halt and the accumulated data
taken as our empirical prior.


\subsection{An Illustration}

An example of a two-stage design involving 16 patients was given by
O'Quigley and Shen (\citeyear{QS1996}).
There were six
levels in the study, maximum likelihood was used, and the first
entered patients were treated at the lowest level. The design was
two-stage. The true toxic probabilities were $R(d_1)=0.03$,
$R(d_2)=0.22$, $R(d_3)=0.45$, $R(d_4)=0.6$, $R(d_5)=0.8$ and
$R(d_6)=0.95$. The working model was that given by
(\ref{powermodel}) where $\alpha_1=0.04$, $\alpha_2=0.07$,
$\alpha_3=0.20$, $\alpha_4=0.35$, $\alpha_5=0.55$ and $\alpha_6=0.70$.
The targeted toxicity was given by $\theta=0.2$ indicating that
the best level for the MTD is given by level 2 where the true
probability of toxicity is 0.22. A grouped design was used until
heterogeneity in toxic responses was observed, patients being
included, as for the classical schemes, in groups of three. The first
three patients experienced no toxicity at level 1. Escalation then
took place to level 2 and the next three patients treated at this
level did not experience any toxicity either. Subsequently two out
of the three patients treated at level 3 experienced toxicity.
Given this heterogeneity in the responses the maximum likelihood
estimator for $a$ now exists and, following a few iterations,
could be seen to be equal to 0.715. We then have that $\hat
R(d_1)=0.101$, $\hat R(d_2)=0.149$, $\hat R(d_3)=0.316$, $\hat
R(d_4)=0.472$, $\hat R(d_5)=0.652$ and $\hat R(d_6)=0.775$. The 10th
entered patient is then treated at level 2 for which $\hat
R(d_2)=0.149$ since, from the available estimates, this is the
closest to the target $\theta=0.2$. The 10th included patient does
not suffer toxic effects and the new maximum likelihood estimator
becomes 0.759. Level 2 remains the level with an estimated
probability of toxicity closest to the target. This same level is
in fact recommended to the remaining patients so that after 16
inclusions the recommended MTD is level 2. The estimated
probability of toxicity at this level is 0.212 and a 90\%
confidence interval for this probability is estimated as (0.07,
0.39).

\section{Large-Sample and Small-Sample Properties}\label{sec.theprop}
Extensive simulations
(O'Quigley, Pepe and Fisher, \citeyear{QPF1990}; O'Quigley and Shen, \citeyear{QS1996};
O'Quigley, \citeyear{Q1999}; Iasonos et al., \citeyear{Ietal2008}), over wide choices of possible true unknown
dose-toxicity situations, show the method to behave in a mostly
satisfactory way, recommending the right level or close levels in
a high percentage of situations and treating in the study itself
a high percentage of included patients, again, at the right level
or levels close by. Cheung (\citeyear{C2005}), O'Quigley (\citeyear{Q2006}) and Lee and Cheung
(\citeyear{LC2009}) obtained theoretical results
which not only provide some confidence in using the method but
can also provide guidance in the choice and structure of working
models.
%
Even though models are misspecified, inference is still based on an estimating
equation taken from the derivative
of the log-likelihood. Thus, Shen and O'Quigley (\citeyear{SQ1996}) defined
\[
I_n(a) = \frac{1}{n} \sum_{j=1}^n  \biggl[ y_j\frac{\psi'}{\psi}
\{ x_j, a \} + (1-y_j) \frac{-\psi'}{1-\psi} \{ x_j, a \}  \biggr].
\]
Some restrictions on $\psi$ are needed (O'Quigley, \citeyear{Q2006}). In particular,
there must exist constants $a_1,\ldots, a_k$
$\in[A,B]$ such that $\psi(d_i,a_i) = R_i $, the function $\psi$
satisfies $\psi(d_i,B) < \theta< \psi(d_i,A)$,
and there is a unique $a_0\in(a_1,\ldots, a_k) $, $\psi(d_0,a_0)
=R(d_0)=\theta_0.$
In general, $\theta_0$ will not be equal to $\theta$ but will be as
close as
we can get given the available doses.
We require the estimating function to respect a standard condition of
estimating functions
which is that
\[
s(t,x,a) = t\frac{\psi'}{\psi}
\{ x, a \} + (1-t) \frac{-\psi'}{1-\psi} \{ x, a \}
\]
is continuous and strictly monotone in $a$. We define
$\tilde{I}_n (a) = n^{-1} \sum_{j=1}^{n} s\{R(x_j),x_j,a \}$.

It is not typically the case that
$\psi(d_i, a_0) = R(d_i)$ for $i = 1, \ldots, k $.
However, at least in the vicinity of the MTD, this will be
approximately true,
an idea that can be formalized (Shen and O'Quigley, \citeyear{SQ1996}) via the
definition of the set
%
\begin{eqnarray}\label{eq:distance-condition}
\qquad S(a_0) &=& \{a\dvtx |\psi(d_0, a) - \theta| < |\psi(d_i , a) - \theta
|\nonumber\\[-8pt]\\[-8pt]
&& \hspace*{92pt}\mbox{for all } d_i \not= d_0 \}.\nonumber
\end{eqnarray}
Shen and O'Quigley (\citeyear{SQ1996}) showed that convergence follows if,
for $i = 1, \ldots, k$, $a_i
\in S(a_0) $.
O'Quigley\break (\citeyear{Q2006}) showed that, for each $1\le i \le k-1$, there exists
a unique constant $\kappa_i$ such that
$\theta-\psi(x_{i},\kappa_{i})=\psi(x_{i+1},\kappa_{i})-\theta> 0.$
The constants $\kappa_i$ naturally give rise to a partitioning of
the parameter space $[A,B].$ 
Letting $\kappa_0 = A$ and $\kappa_k = B$, we can write the
interval $[A,B]$ as a union of nonoverlapping intervals whereby
$[A,B] = \bigcup_{i=1}^{k} [\kappa_{i-1}, \kappa_i) .$
This partition is of particular value in establishing prior
distributions which
can translate immediately into priors for the dose levels themselves.
It is
also of value in deriving results concerning the coherence, stability
and convergence
of the algorithm (Cheung and Chappell, \citeyear{CC2002}; O'Quigley, \citeyear{Q2006}).
Convergence to the MTD stems from the fact that $\sup_{a\in[A,B]}
|I_n(a) - \tilde{I}_n(a) | $
converges almost surely to zero (Shen and O'Quigley, \citeyear{SQ1996}) and that we
can re-express
%
$\tilde{I}_n(a) $ as a sum over the $k$ dose levels rather than a sum over
the $n$ subjects; in particular we have that $\tilde{I}_n(a)= \sum
_{i=1}^k \pi_n(d_i) s\{R(d_i),d_i,a \} .$
Supposing that the solution to the equation $\tilde{I}_n(a)= 0$ is
$\tilde{a}_n$
and that $a_i$ is
the unique solution to the equation $s\{R(d_i),d_i,a \} = 0$,
then $\tilde{a}_n$ will fall into the interval
$S_1(a_0)$. Since $\hat{a}_n$ solves $I_n(a) =0$,
then, almost surely,
$\hat{a}_n \in S(a_0),$ so that, for $n$ sufficiently large, $x_{n+1}
\equiv d_0.$
Since there are only a finite number of dose levels, $x_n$
will ultimately settle at $d_0.$ Rather than appeal to the set
$S(a_0),$ which
quantifies the roughness of the working approximation to the true
dose-toxicity function
in the vicinity of the MTD, and which guarantees convergence to the MTD
when all of the
$a_i$ belong to this set, Cheung (\citeyear{C2005}) used a related approach which
appeals to a more
flexible---in many ways more realistic---definition of the MTD whereby
probabilities of toxicity
within some given range are all taken to be acceptable. Convergence can
then be shown to obtain
without such restrictive conditions as those described
above.\looseness=-1

\subsection{Efficiency}\label{sec4.1}
O'Quigley (\citeyear{Q1992}) proposed using $\hat{\theta}_n= \psi(x_{n+1},
\hat{a}_n)$ to estimate the probability of toxicity at the
recommended level $x_{n+1}$, where $\hat{a}_n$ is the maximum
likelihood estimate. An application of the $\delta$-method (Shen
and O'Quigley, \citeyear{SQ1996}) shows that the asymptotic distribution of
$\sqrt{n}\{\hat{\theta}_n- R(d_0)\}$ is
$N\{0,\theta_0(1-\theta_0)\}$. The estimate then provided by CRM
is fully efficient for large samples. This is what our intuition
would suggest given the convergence properties of CRM. What
actually takes place in finite samples needs to be investigated
on a case by case basis. The relatively broad range of cases
studied by O'Quigley (\citeyear{Q1992}) show a mean squared error for the
estimated probability of toxicity at the recommended level under
CRM to correspond well with the theoretical variance for samples
of size $n$, were all subjects to be experimented at the correct
level. Some of the cases studied showed evidence of
super-efficiency, translating nonnegligible bias that happens to
be in the right direction, while a few others indicated efficiency
losses large enough to suggest the possibility of better performance.

A useful tool in studies of finite sample efficiency is the idea
of an optimal design. We can derive a nonparametric optimal
design (O'Quigley, Paoletti and \mbox{Maccario}, \citeyear{QPM2002}) based upon no more
than a
monotonicity assumption. Such an optimal design is not
generally available in practice but can serve as a gold standard
in theoretical studies, playing a role similar to that of the
Cramer--Rao bound.
Comparisons between any suggested method and the optimal
design enable us to quantify just how much room there is for potential
improvement.
Further evidence of the efficiency of the CRM was provided by the work
of O'Quigley, Paoletti and Maccario (\citeyear{QPM2002}), where the CRM
is compared to the nonparametric optimal design. In the cases studied
in that article and in that of Paoletti, O'Quigley and Maccario
(\citeyear{PQM2004}), potential for improvement is seen to be limited, with the
identification of the MTD by the two-stage CRM design being only
slightly inferior to that of the optimal design.
\subsection{Nonidentifiability of Fully Parameterized
Models}\label{ssec.nonid}

Under the conditions outlined above we
will ultimately only include patients at dose level $d_0$. Under
very much broader conditions (Shen and O'Quigley, \citeyear{SQ1996}) we can
guarantee convergence to some level, not necessarily $d_0$ but
one where the probability of toxicity will not be far removed
from that at $d_0$. The consequence of this is that, for the most
common case of a single homogeneous group of patients, we are
obliged to work with an underparameterized model, notably a
one-parameter model in the case of a single group. Although a
two-parameter model may appear more flexible, the convergence
property of CRM means that ultimately we will not obtain
information needed to fit two parameters. Having settled at dose
level $d_i$, the only quantity we can estimate is $R(d_i)$ which
can be done consistently in light of the Glivenko--Cantelli lemma.
Under our model conditions we have that $R(d_i)=\psi(d_i,a_i)$
and that ${\hat a}_j$ will converge almost surely to $a_i$.
Adding a second parameter can only overparameterize the situation
and, for example, the commonly used logistic model has an
infinite number of combinations of the two parameters which lead
to the same value of $R(d_i)$. A likelihood procedure can then be
unstable and may even break down, whereas a two-parameter fully
Bayesian approach (Gatsonis and Greenhouse, \citeyear{GG1992}; Whitehead and
Williamson, \citeyear{WW1998}) may work initially, although somewhat
artificially, but behave erratically as sample size increases and
the structural rigidity provided by the prior gradually wanes.
This is true even when starting out at a low or the lowest level,
initially working with an Up and Down design for early
escalation, before a CRM model is applied. Indeed, any design
that ultimately concentrates all patients from a single group on
some given level can fit no more than a single parameter without
running into problems of identifiability.

\section{Extended CRM Designs}\label{sec5}
The simple model of (\ref{powermodel}) can be extended to a class of
models denoted
by $\psi_m (x_j,a)$ for $m=1, \ldots, M$
where there are $M$ members of the class. Take, for example,
%
\begin{eqnarray}\label{powermodels}
\psi_m (d_i,a) =\alpha_{mi}^{\exp(a) },  \nonumber\\[-8pt]\\[-8pt]
\eqntext{i=1,\ldots,k; m=1,\ldots, M,}
\end{eqnarray}
where $0<\alpha_{m1}< \cdots<\alpha_{mk}<1$ and $-\infty<a<\infty
$, as an immediate generalization of
(\ref{powermodel}).
Prior information concerning the plausibility of each model is catered
for by $\pi(m)  $, $m=1, \ldots,M,$ where
$\pi(m) \geq0$ and where $\sum_m \pi(m) =1.$ When each model is
given the same initial weighting, then we
have that $\pi(m) = 1/m .$ If the data are to be analyzed under model
$m$, then, after the inclusion of $j$ patients, the
logarithm of the likelihood can be written as
\begin{eqnarray}
{\mathcal{L}}_{mj}(a) & = & \sum_{\ell=1}^{j} y_\ell  \log
\psi_m (x_\ell,a) \nonumber\\[-8pt]\\[-8pt]
&&{}+ \sum_{\ell=1}^{j} (1-y_\ell)
\log\bigl(1-\psi_m (x_\ell,a)\bigr),\nonumber
\end{eqnarray}
where any terms not involving the parameter $a$ have been ignored.
Under model $m$ we obtain a summary value of the parameter $a$, in
particular the maximum of the posterior
mode and we refer to this as ${\hat a}_{mj}$. Given the value of ${\hat
a}_{mj}$ under model $m$,
we have an estimate of the probability of
toxicity at each dose level $d_i$ via $ {\hat
R}(d_i)=\psi_m(d_i,\hat{a}_{mj})   $ $  (i=1,\ldots,k). $
On the basis of this formula, and having taken some value for $m$, the
dose to be given to the
$(j+1) $th patient, $x_{j+1}$, is determined. Thus, we need some value
for $m$ and we make use of the
posterior probabilities of the models given the data $\Omega_j.$
Denoting these posterior probabilities by $\pi(m|\Omega_j)$, then
\begin{eqnarray}
\quad &&\pi(m|\Omega_j)\nonumber\\[-8pt]\\[-8pt]
\quad &&\quad  = \frac{ \pi(m)\int_{-\infty
}^\infty\exp\{ {\mathcal{L}}_{mj}(u)\} g(u)  \,du }{ \sum_{m=1}^M \pi
(m)\int_{-\infty }^\infty\exp\{ {\mathcal{L}}_{mj}(u)\} g(u)  \,du }.\nonumber
\end{eqnarray}
The estimated values of $\pi(m|\Omega_j)$ can help us decide between
models which have physical implications of
interest to us. As an example suppose that there exists significant
heterogeneity among the patients and this is simplified
to the case of two groups.

\subsection{A Simple Heterogeneity Model}\label{sec5.1}
As in other types of clinical trials we are essentially looking
for an average effect. Patients naturally differ in the way they
may react to a treatment and, although hampered by small samples,
we may sometimes be in a position to specifically address the
issue of patient heterogeneity. One example occurs in patients
with acute leukemia where it has been observed that children will
better tolerate more aggressive doses (standardized by their
weight) than adults. Likewise, heavily pretreated patients are
more likely to suffer from toxic side effects than lightly
pretreated patients. In such situations we may wish to carry out
separate trials for the different groups in order to identify the
appropriate MTD for each group. Otherwise we run the risk of
recommending an ``average'' compromise dose level, too toxic for
a part of the population and suboptimal for the other. Usually,
clinicians carry out two separate trials or split a trial into
two arms after encountering the first DLTs when it is believed
that there are two distinct prognostic groups. This has the
disadvantage of failing to utilize information common to both
groups. The most common situation is that of two samples where we
aim to carry out a single trial keeping in mind potential
differences between the two groups. A multisample CRM is a
direct generalization although we must remain realistic in terms
of what is achievable in the light of the available sample sizes.

Introduce a binary variable $Z$ taking the value 0 for the first group
and 1 for the second group. Suppose
also that we know that, for the second group, the probability of
toxicity can only be the same or higher than the first
group. For this situation consider the following models:\looseness=-1
\begin{enumerate}
\item{Model 1: $m=1$}
\begin{eqnarray*}\label{model2sga}
&& \Pr(Y=1 | d_i, z=0) = \psi
(d_i,a)   , \quad  i=1,\ldots, k ,\\
&& \Pr(Y=1 | d_i, z=1) = \psi(d_i,a)  ,\quad  i=1,\ldots, k ,
\end{eqnarray*}
\item{Model 2: $m=2$}
%
\begin{eqnarray}\label{model2sga2}
&& \Pr(Y=1 | d_i, z=0) = \psi
(d_i,a)   ,\quad  i=1,\ldots, k ,\nonumber \\
&& \Pr(Y=1 | d_i, z=1) =
\psi(d_{i+1},a)  ,\nonumber\\
\eqntext{i=1,\ldots, k-1,} \\
&& \Pr(Y=1 | d_i,
z=1) = \psi(d_{k},a)  ,\quad  i= k.\nonumber
\end{eqnarray}
\end{enumerate}

If the most plausible model has $m=1$, then we conclude that there is
no difference between the groups. If $m=2$,
then we conclude that for the second group the probability of toxicity
at any level is the same as that for a
subject from the first group but treated at one level higher. The truth
will be more subtle but since we have to
treat at some level we force this decision to be made at the modeling
stage. The idea extends, of course, to several
levels, positive as well as negative directions to the difference, and
to other factors such as treatment schedules.

\subsection{Randomization and Two-Parameter Models}\label{sec5.2}

Suppose that $j$ subjects are already entered in the trial.
Instead of systematically selecting the level estimated as being
closest to the target, a different approach would be to use the
available knowledge to randomly select a level from $d_1, \ldots,
d_k$ according to some given discrete distribution. This
distribution does not have to be fixed in advance but can depend
on the available levels and the current estimate of the MTD.
Let $x_{j+1}$ be defined as before. However, we will no longer
allocate systematically subject $j+1$ to dose level $x_{j+1}$ as
before. Instead we allocate to $w_{j+1}$ where we define
%
\begin{equation}
\quad w_{j+1} =
\left\{
\begin{array}{l}
\displaystyle\sum_{m=1}^k d_{m+\Delta} I\{x_{j+1}=d_m ,  m<k \}  ;\vspace*{2pt}\cr
\quad {\hat R}(x_{j+1})\leq\theta \vspace*{3pt}\cr
\displaystyle\sum_{m=1}^k d_{m-\Delta} I\{x_{j+1}=d_m ,  m>1  \}  ;\vspace*{2pt}\cr
\quad {\hat R}(x_{j+1})>\theta
\end{array}
\right\}
\end{equation}
%
and where $\Delta$ is a $\operatorname{Bernoulli} (0,1)$ random variable with
parameter typically of value 0.5. In words, instead of allocating to the
level closest to ${\hat R}(x_{j+1})$ we allocate, on the basis of
a random mechanism, to the level just above ${\hat R}(x_{j+1})$
or the level just below ${\hat R}(x_{j+1}).$ In the cases where
${\hat R}(x_{j+1})$ is lower than the lowest available level, or
higher than the highest available level, then the allocation
becomes, again, systematic. The purpose of the design is then to
be able to sample on either side of the target. Aside from those
cases in which the lowest level appears to be more toxic than the
target or the highest level less toxic than the target,
observations will tend to be concentrated on two levels. One of
these levels will have an associated estimated probability below
the target while the other level will have an estimated
probability above the target. 

An immediate consequence of forcing experimentation to take place
at more than a single level is that the nonidentifiability
described above changes. It is now possible to estimate more than
a single parameter, for example the rate of toxicity at, say, the
lower of the two levels as well as the rate of toxicity at the
next level up. Working with a one-parameter model and randomizing
to two levels, say $d_\ell$ and $d_{\ell+1}$, the estimate $\hat
a$ will converge to the solution of the equation
\[
{\pi}(d_\ell)
 \biggl\{R_\ell\frac{\psi'}{\psi}
(d_\ell,a) + (1-R_\ell) \frac{-\psi'}{1-\psi}(d_\ell,a)
 \biggr\},\vspace*{-5pt}
\]
\begin{eqnarray*}
&&\{1-{\pi}(d_\ell)\}  \biggl\{R_{\ell+1} \frac{\psi'}{\psi}
(d_{\ell+1},a)\\
&&\hphantom{\{1-{\pi}(d_\ell)\}  \biggl\{}
{}+(1-R_{\ell+1}) \frac{-\psi'}{1-\psi}(d_{\ell+1},a)
 \biggr\}=0,
\end{eqnarray*}
where $\pi(d_\ell)$ is the stable distribution (long-term
proportion) of patients included at level $d_\ell$. Comparing
this equation with the estimating equation for the standard case without
randomization, we can see
that, unless the working model generates the
observations, we will not obtain consistent estimates of the
probabilities of
toxicities at the two doses of the stable distribution. However,
introducing a second parameter into the model, one which
describes the differences between the probabilities of toxicity
at the two dose levels, we obtain consistent estimates at these
two doses of the stable distribution. To see this it is enough to
parameterize the probability of toxicity at the current level
$d_\ell$ as $\psi(d_\ell,a)$ and that at level $d_{\ell+1}$ by
$\psi(d_\ell,a+b).$ The estimates will converge to the solution of
\[
{\pi}(d_\ell)
 \biggl\{R_\ell\frac{\psi'}{\psi}
(d_\ell,a) + (1-R_\ell) \frac{-\psi'}{1-\psi}(d_\ell,a)
 \biggr\},\vspace*{-5pt}
 \]
\begin{eqnarray*}
\\ & & \{1-{\pi}(d_\ell)\}  \biggl\{R_{\ell+1} \frac{\psi'}{\psi}
(d_{\ell+1},a+b)\\
&&\hphantom{ \{1-{\pi}(d_\ell)\}  \biggl\{}
{}+(1-R_{\ell+1}) \frac{-\psi'}{1-\psi}(d_{\ell+1},a+b)
 \biggr\}=0,
\end{eqnarray*}
for which each term separately can be then accommodated within
the framework describing consistency given above.
%
In practice we would use a model such as the logistic where
%
%
\[
\psi(d_k,a,b)=\frac{\exp(a \alpha_k+b)}{1+\exp(a \alpha_k+b)},
\]
which, once settling takes place, is then a saturated model.

\section{Related Designs}\label{sec6}
There have been many suggestions in the literature for possible
modifications of the basic
design. Also, some apparently alternative designs turn out to be
equivalent to the basic
design. In this section we consider some of these designs.

\subsection{Escalation with Underdose/Overdose Control}\label{sec6.1}
Babb, Rogatko and Zacks (\citeyear{BRZ1998}) argued that the main ethical concern
was not so much
putting each patient at a dose estimated to be the closest to the MTD
but rather putting
each patient at a dose for which the probability of it being too great
was minimized. The
difference may be subtle but would be a basis for useful, and
important, discussions with
the clinicians involved. These discussions help make explicit the
goals, both in terms of
final recommendation and for those patients included in the study.
There may be situations
where a parallel concern might focus on the underdosing rather than the
overdosing.
For an approach based on the CRM we would simply modify the definition
of the dose level ``closest to the target'' to be asymmetric. Positive
distances could be magnified relative to negative ones resulting in a
tendency to assign below the MTD rather than above it.

Babb, Rogatko and Zacks (\citeyear{BRZ1998}) approached the problem differently by
focusing on the posterior
distribution of the MTD and suggesting a loss function that penalizes
overdosing to a greater degree than underdosing. Tighiouart, Rogatko and Babb (\citeyear{TRB2005}) developed the idea further, investigating a number of
prior distributions. Despite this change in emphasis, there is no
fundamental difference between these approaches and the CRM, aside from
the making use of a particular distance measure. The methods of Babb, Rogatko and Zacks (\citeyear{BRZ1998}) and Tighiouart, Rogatko and Babb (\citeyear{TRB2005})
allow for continuous dose levels. Although the CRM is most frequently
applied in cases with a fixed set of dose levels, it can be adapted to
allocate patients on dose levels other than the fixed set of doses.


\subsection{ADEPT and Two-Parameter CRM}\label{sec6.2}
O'Quigley, Pepe and Fisher (\citeyear{QPF1990}) studied two-parameter
CRM models based on the logistic
distribution. For large samples the parameters are not identifiable and
we may expect that
this could lead to unstable or undesirable operating characteristics.
For small to moderate finite
samples the behavior can be studied on a case by case basis. Even when
the two-parameter model
correctly generated the observations, the simulations of O'Quigley,
Pepe and Fisher indicated that
the one-parameter CRM would work better for sample sizes up to around 25.

Whitehead and Brunier (\citeyear{WB1995}) suggested working with the two-parameter
logistic model and using
a pseudo-data prior. This has been put together as a software package
and is called ADEPT. The
term ADEPT is used to describe either the software itself or the
approach which would be equivalent
to a two-parameter CRM with a data-based prior. Gerke and Siedentop
(\citeyear{GS2008}) argued that
ADEPT is to be preferred to standard CRM in terms of accuracy of
recommendation. This conclusion
was based on a study of three, rather particular, situations in which
the target dose lies exactly at the midpoint between two of the available
doses. They define the lower of these two doses as being the MTD. Gerke
and Siedentop's definition of the MTD is
not the usual one which, had it been used in their simulations, would
have resulted in the very
opposite conclusion. The usual one, and that used in O'Quigley, Pepe
and Fisher, is the dose
which is the closest to the target. Should two doses be equidistant
from the target then, logically,
either one could be considered to be the MTD. This observation alone
would completely reverse the
findings of Gerke and Siedentop (Shu and O'Quigley, \citeyear{SQ2008}).

The ADEPT program leans more formally on\break Bayes\-ian decision procedures
which, it is argued\break (Whitehead and Brunier, \citeyear{WB1995}),
represent a generalization of the CRM since, instead of basing
sequential patient
allocation on the anticipated gain for the next included patient or
group of patients, allocation could be
based on the gain
for the variance of estimators. In the case of more than one parameter
we could use different
combinations of the individual variances and
covariances, in particular the determinant of the information matrix.
Whitehead and
Brunier argued that ``gain functions can be devised from the point of view
of the investigator (accuracy for future patients) or from the point of
view of the next
included patient, as in the CRM. Weighted averages of
these two possibilities can be used to form compromise procedures.''

However, under current guidelines, it is not possible to
use a procedure which sacrifices the point of view of the current patient
to that of future patients. It is only future patients who may benefit
from improved precision (the
point of view of the investigator) and, although, in medical experimentation,
arguments have been and will continue to be put
in such a direction, such logic is not currently considered acceptable.
Experimentation
on an individual patient can only be justified if it can be argued that
the driving goal is
the benefit of that same patient.
Basing allocation
on anything other than patient gain, and, in particular, the gain for
future patients,
would be a violation of the usual ethical criteria in force
in this area. In practice, only patient gain is used, and so ADEPT is
essentially the same as
two-parameter CRM. In animal experimentation or in experimentation in
healthy volunteers, where severe
side effects are considered very unlikely, a case could be built for
using other gain functions.

\subsection{Curve-Free Designs}\label{sec6.3}
Rather than appeal to a working model $\psi(x,a)$ and have $a$
follow some distribution, we can employ a multivariate
distribution of dimension $k$ and consider the ordered
probabilities at the $k$ levels to be the quantities of interest.
Prior median or mean values for the distribution of $R(d_i)$, the
probability of toxicity at dose $d_i$, are provided by the
clinician. We then work with a multivariate law that is flexible
enough to allow reasonable operating characteristics, escalating
quickly enough in the absence of observed toxicities and not
being unstable or overreacting to toxicities that occur.
Gasparini and Eisele (\citeyear{GE2000}) argued in favor of experimenting this
way. They suggested working with a product of beta priors (PBP) upon
reparameterizing whereby
\begin{eqnarray*}
\theta_1&=& 1-R(d_1)  , \\
\theta_i &=& \frac
{1-R(d_i)}{1-R(d_{i-1})}  ,\quad
i= 2, \ldots, k,
\end{eqnarray*}
and then letting the $\theta_i $ $ (i=1,\ldots,k)$ have
independent beta distributions. Since $R(d_i)= 1-\theta_1\theta_2
\cdots \theta_i $ the monotonicity constraint is respected. The
distribution of a product of beta distributions is complex but
the authors argue that we can approximate this well by taking the
product itself to be beta. We then fit such a beta using the
first two moments from the product, easily achieved under the
condition of independence of the $\theta_i$. Gasparini and Eisele
(\citeyear{GE2000}) provided some guidelines for setting up the prior for this
multivariate law based on consideration of operating
characteristics. O'Quigley (\citeyear{Q2002b}) demonstrated an equivalence between
a curve-free design and a CRM design in that, given a particular specification
of a curve-free design, there exists an equivalent specification of a
CRM design. This
is also true in the other direction. By equivalent
we mean that all operational characteristics are the same. However,
this still
remains only an existence result and it is not yet known how to
actually find the
equivalent designs. Cheung (\citeyear{C2002}) noted that in cases where low
toxicity percentiles are targeted, the use of the nonparametric
approach with a vague prior can lead to dose allocation that tends to
be confined to suboptimal levels. Cheung (\citeyear{C2002}) exploited the
connection with the CRM to suggest informative priors that can help
alleviate this problem.

Whitehead et al. (\citeyear{WTW2010}) suggested an approach in which the
probabilities of toxicity at each dose, rather than
belonging to some continuum, are only allowed to belong to a small
discrete set. In practice, we do not
need to distinguish a probability of toxicity of 0.32 from a probability
of 0.34. They could be considered the
same, or, in some sense at least, equivalent. The idea is not unrelated
to the idea of Cheung and Chappell (\citeyear{CC2002}) on
indifference intervals. Performance of Whitehead and colleague's method
is comparable to the CRM.

\section{Identifying the Most Successful Dose (MSD)}\label{sec7}
In the context of dose finding in HIV, O'Quigley, Hughes and Fenton (\citeyear{QHF2001})
considered the problem of finding the dose which maximizes the overall
probability of
success. Here, failure is either a toxicity (in the HIV context, mostly
an inability to
maintain treatment) or an unacceptably low therapeutic response. Zohar
and O'Quigley (\citeyear{ZQ2006a})
made a slight modification to the approach to better accommodate the
cancer setting.
We take $Y$ and $V$
to be binary random variables $(0,1)$ where $Y=1$ denotes a toxicity,
$Y=0$ a nontoxicity, $V=1$ a response, and $V=0$ a nonresponse.
As before, the probability of
toxicity at the dose level $X_j=x_j$ is defined by
\[
R(x_j)=\Pr(Y_j=1|X_j=x_j).
\]
The probability of response
given no toxicity at dose level $X_j=x_j$ is defined by
\[
Q(x_j)=\Pr(V_j=1|X_j=x_j,Y_j=0),
\]
so that $P(d_i)=Q(d_i)\{1-R(d_i)\}$ is the probability of success.
A successful trial would identify the dose level $l$ such
that $P(d_l)>P(d_i)$ (for all $i$ where $i\neq l$). Zohar and O'Quigley (\citeyear{ZQ2006b})
called this
dose the most successful dose and our purpose in this
kind of study is, rather than find the MTD, to find the MSD.
The relationship between toxicity and dose ($x_j$) and the
relationship between response given no toxicity and dose can be
modeled through the use of two one-parameter models.
Whereas $R(d_i)$ and $Q(d_i)$ refer to exact, usually unknown,
probabilities, the model-based equivalents of these, $\psi$
and $\phi$, respectively, are only working approximations given by
\begin{eqnarray*}
R(d_i)&\approx&\psi(d_i,a)=\alpha^{\exp a}_{i} ;\\
Q(d_i)&\approx&\phi(d_i,b)=\beta^{\exp b}_{i},
\end{eqnarray*}
where
$0<\alpha_1<\cdots<\alpha_k<1$, $-\infty<a<\infty$,
$0<\beta_1<\cdots<\beta_k<1$ and $-\infty<b<\infty$. For each
dose, there
exist unique values of $a$ and $b$ such that the approximation becomes
an equality at that dose,
but not necessarily exact at the other doses. After the inclusion
of $j$ patients, $R(d_i)$, $Q(d_i)$, and $P(d_i)$ are estimated by
\begin{eqnarray*}
\hat{R}(d_i)&=&\psi(d_i,\hat{a}_j)  ;\qquad
\hat{Q}(d_i)=\phi(d_i,\hat{b}_j)  ;\\
\hat{P}&=&\phi(d_i,\hat{b}_j)\{1-\psi(d_i,\hat{a}_j)\},
\end{eqnarray*}
where
$\hat{a}_j$ and $\hat{b}_j$ maximize the log-likelihood (see
O'Quigley, Hughes and Fenton, \citeyear{QHF2001}).

\section{Conclusions}\label{sec8}
More fully Bayesian approaches in a decision making context,
and not simply making use of Bayesian estimators,
have been suggested for use in the context of Phase I trial
designs. These can be more in the Bayesian spirit of
inference, in which we quantify prior information, observed from
outside the trial as well as that solicited from clinicians
and/or pharmacologists. Decisions are made more formally using
tools from decision theory. Any prior information can subsequently
be incorporated via the Bayes formula into a posterior density
that also involves the actual current observations. Given the
typically small sample sizes often used, a fully Bayesian approach
has some appeal in that we would not wish to waste any relevant
information at hand. Unlike the set-up described by O'Quigley, Pepe and
Fisher (\citeyear{QPF1990}), we could also work with informative priors.

Gatsonis and Greenhouse (\citeyear{GG1992}) considered two-parameter probit and
logit models for dose response and studied the effect of different
prior distributions. Whitehead and Williamson (\citeyear{WW1998}) carried out
similar studies but with attention focusing on logistic models
and beta priors. Whitehead and Williamson (\citeyear{WW1998}) worked with some
of the more classical notions from optimal design for choosing the
dose levels in a bid to establish whether much is lost by using
suboptimal designs. O'Quigley, Pepe and Fisher (\citeyear{QPF1990})
ruled out
criteria based on optimal design due to the ethical criterion of
the need to attempt to assign the sequentially included patients
at the most appropriate level for the patient. This same point was
also emphasized by \mbox{Whitehead} and Williamson (\citeyear{WW1998}). Certain contexts,
however, may allow the use of more formal optimal procedures.

For certain problems we may have good knowledge about some aspect of
the problem
and poor knowledge on the others. The overall dose-toxicity curve may
be very poorly
known but, if this were to be given for, say, one group, then we would
have quite strong knowledge
of the dose-toxicity curve for another group.
Uninformative Bayes or maximum likelihood would then seem
appropriate overall although we would still like to use information
that we have, an
example being the case of a group weakened by extensive prior therapy
and thereby very
likely to have a level strictly less than that for the other
group. Careful parameterization would enable this information to be
included as a constraint. However, rather than work with a rigid
and unmodifiable constraint, a Bayesian approach would allow us to
specify the anticipated direction with high probability while
enabling the accumulating data to override this assumed direction
if the two run into serious conflict. Exactly the same idea could
be used in a case where we believe there may be group
heterogeneity but that it be very unlikely the correct MTDs differ
by more than a single level. This is especially likely to be of relevance
in situations where a defining prognostic variable, say the amount of
prior treatment, is not very sharp so that group classifications may be
subject to some error. If the resulting MTDs do differ we would not
expect the
difference to be very great. Incorporating such information into the
design will
improve efficiency.

Stochastic approximation, which is an algorithm for finding the
root of an unknown regression equation, can be shown, under
certain conditions, to be equivalent to recursive inversion of a
linear model (Wu, \citeyear{W1985}, \citeyear{W1986}; Cheung and Elkind, \citeyear{CE2010}). In the light of
those results, the
CRM, in its basic form, could then be
viewed as stochastic approximation leaning upon a particular
dose-response model rather than a linear one. However, this
characterization of the methodology is less fundamental than two
others: (1) use of an underparameterized model and (2)
restriction of the available doses to a limited finite set.

The
second of the above characterizations implies the necessity for the
first (see Section \ref{ssec.nonid}). Consistency of stochastic
approximation fails in the setting where we have a limited set of
available responses (doses) and can only be achieved under
conditions analogous to those outlined in this article (Shen and
O'Quigley, \citeyear{SQ2000}). Other algorithms similar to stochastic approximation (adaptive
designs) rely on probabilistic rules to identify some percentile
(dose) from an unknown distribution. Wu's (\citeyear{W1985}, \citeyear{W1986}) findings
suggest that there is usually some implicit model behind the
algorithm.

The CRM makes implicit
models explicit ones; underparameterized, and therefore
misspecified, but sufficiently flexible to obtain accurate
estimates locally although not reliable at points removed from
those at which the bulk of experimentation takes place. The
model, being explicit, readily enables extension and
generalization. The two group case, incorporation of
randomization about the target or the inclusion of partial prior
information are, at least conceptually, relatively
straightforward tasks. The framework is then in place to
investigate other aspects of dose-finding designs such as
multigrade outcomes or the ability to exploit information on
within-subject escalation. As for any method, there is always room for
improvement,
although the results on optimality suggest that, for the basic problem,
this room is
not great. It is likely to be more fruitful to focus our attention on
more involved problems
such as continuous outcomes, subject heterogeneity, combined
efficacy-toxicity studies, and
studies involving escalation of two or more components.

%

\section*{Acknowledgments}
The authors acknowledge the input of the reviewers and editors,
in particular for suggestions which have improved the clarity and for
bringing to our
attention relevant material that we had overlooked. This work was
supported in part by research Grant 1R01CA142859-01A1. Designs for
phase I trials of combinations of agents.

\vspace*{-3pt}

\end{document}